# A unified approach to the thermodynamics of a photovoltaic system

Ido Frenkel & Avi Niv*

Department of Solar Energy and Environmental Physics, Jacob Blaustein Institutes for Desert Research, Ben-Gurion University of the Negev, Sede Boqer Campus, 8499000, Israel

* Corresponding author: aviniv@bgu.ac.il

## Summary Paragraph

Thermodynamics is accepted as a universal truth, encompassing all macroscopic objects. Therefore, it is surprising to find that, within our current understanding, the photovoltaic effect has so far eluded the first and second laws of thermodynamics. The inconsistency emerges from the fact that photovoltaics obey a distinct law of detailed balance[1]. Since radiative processes depend on only two independent variables that are the chemical potential and the temperature, the detailed balance and the two laws of thermodynamics cannot be mutually solved. In this work, we resolve this incompatability by proposing that the system is controlled by yet a third independent variable, which is related to the emissivity. This unification not only advances our fundamental understanding of light-matter interactions but, perhaps more importantly, allows us to assess the limiting factors of advanced photovoltaic concepts designed for elevated temperatures. These include thermophotovoltaics[2], thermoradiative[3–5] and thermophotonic solar power conversion[6], and radiative cooling[7–10], which are instrumental in our ability to develop advanced renewable energy technologies.

## Introduction

In their seminal paper, Shockley and Queisser (S&Q) established the limiting efficiency of a photovoltaic (PV) solar cell based on the equality in the number of excited and depleted electron-hole pairs, a principle known as detailed balance (DB)[1]. Alternatively, there is also a thermodynamic estimation of the maximal work that is available from the sun, which is based on energy balance (EB) and entropy balance (SB)—the first and second laws of thermodynamics[13–15]. The two approaches are not the same, however. While the thermodynamic estimation is at about 94%, the detailed balance efficiency limit is a little over 30% for a single junction cell. Given that thermodynamics is universal, an obvious question arises: How does a DB-maintaining system, such as a photovoltaic (PV) cell, fulfill the thermodynamic requirements of EB and SB?

Surprisingly, this question is still open since, in their present form, the three laws that supposedly govern the PV effect, namely the detailed, energy, and entropy balances, form an overdetermined set of constraints. This inconsistency emerges since, within each of these three rules, radiative processes are a function of only two independent variables: the temperature and the (chemical) potential. Past attempts to resolve this inconsistency were based on finding a thermodynamic justification for the DB law, thus eliminating it as an independent constraint[16–19]. Despite the theoretical progress achieved in this manner[20,21], a full thermodynamic description of the PV effect has not emerged. It is our view that the DB is a law in its own right, which exists in addition to the thermodynamic EB and SB laws, rather than emanating from them.

In this paper, we solve this inconsistency by proposing that radiative processes are a function of a third independent variable, rather than just the former two. The choice of this third variable is motivated by the fact that the amount of radiation is always proportional to the property of *emissivity*, and also by observing the temperature-dependent emissivity of semiconductors[22–26].

We, therefore, propose that there has to be a background emissivity that acts as an independent variable, rather than being a surface or an electronic-structure-dependent property of a material. Reformulating the detailed, energy, and entropy balances accordingly, the abovementioned inconsistency dissolves, and a fully consistent set of constraints that apply to any material emerges. To distinguish this independent variable from what is usually referred to as emissivity, we call it the *thermodynamic emissivity* (TE) and label it $\varepsilon_T$.

We follow by demonstrating our unified approach in the important test case of a solar cell. Specifically, we analyze a fully isolated solar cell, a cell that is electrically isolated but thermally conductive, and a cell that performs electrical work while conducting heat to its environment. We show that for infinite heat conduction, the proposed model recedes to the well-known S&Q limit.[1] Solving the thermodynamic puzzle of the PV effect not only gives us a better understanding of light-matter interactions but also aids in the utilization of renewable energy sources, given the central role of PV in this process[2,3,12,4–11].

## The inconsistency of present approaches

S&Q calculated the efficiency of a single junction solar cell by assuming that each absorbed photon creates a single electron-hole pair and that the number of generated and depleted electron-holes is equal[1]. The equality in generation and depletion rates is called a *detailed balance* (DB). In the absence of non-radiative recombinations, the DB is stated as follows:

$$N_s = N_c(V, T) + I, \qquad (1)$$

where $N_s$, is the rate of photons coming from the source, $N_c(V,T)$ is the radiative recombination rate at potential $V$ and temperature $T$, and $I$ is the electric current.

S&Q solved Eq. (1) to find the work from the cell, which is the product $V \cdot I$, as a function of the bandgap by using the sun as a source. The problem is that both potential and temperature enter the radiative recombination rate term $N_c(V,T)$. Therefore, Eq. (1) alone does not suffice, and the S&Q analysis must be supplemented with an arbitrary nominal temperature.

Assuming a fixed temperature, however, means that there must be a constant energy flux emanating from the system to its surroundings. According to the first law of thermodynamics, this energy flux is determined according to the following energy balance (EB) law:

$$E_s = E_c(V,T) + E_{eh}I + Q, \qquad (2)$$

where $E_s$ is the energy flux from the source, $E_c(V,T)$ is the radiation energy flux from the system, $E_{eh}$ is the energy that a single electron-hole pair carries out of the system as an electrical current, and $Q$ is the heat flux conducted to the environment.

There is also the second law of thermodynamics that has to be fulfilled, stated here as an entropy balance (SB) law:

$$S_s + S_g = S_c(V,T) + S_{eh}I + \frac{Q}{T}, \qquad (3)$$

where $S_s$ is the entropy flux received from the source, $S_c(V,T)$ is the entropy flux leaving the system due to radiative recombinations, and $S_{eh}$ is the entropy of an electron-hole pair. Here, however, there is an additional entropy flux $S_g > 0$, which is the entropy generated in the process of converting a photon to an electron-hole pair. An essential characteristic of the energy and entropy equations is that minimizing the generated entropy flux $S_g$ maximizes the work. Therefore, the constraint that the second law of thermodynamics imposes on the system is the minimization of $S_g$.

Equations (2) and **Error! Reference source not found.** are completely universal, regardless of whether the system is radiative or not[13–15]. An example of a simple conductive system that obeys these two rules is the Carnot heat engine: Taking $E_c = S_c = 0$, $E_{eh}I = W$, $S_{eh}I = 0$, and $T_s$ for the source temperature, and then canceling Q between Eqs. (2) and **Error! Reference source not found.**, gives:

$$\frac{W}{E_s} = \left(1 - \frac{T}{T_s}\right) + \frac{TS_g}{E_s}.$$

The above expression is the Carnot efficiency at the limit of $S_g \to 0$. Likewise, for the radiative case, the Landsberg efficiency limit is obtained by inserting the proper terms for the energy and entropy fluxes of blackbody radiaiton and taking the limit of $S_g \to 0$ as before[13,27].

Returning to the matters at hand, subjecting a PV system to the governing laws of Eqs. (1)–**Error! Reference source not found.** results in an inconsistency since three constraints are placed on a two-variable system, which is $T$ and $V$. This inconsistency leads to the following conclusion: *If a system is subjected to the DB and the two laws of thermodynamics, then in addition to the temperature $T$ and potential $V$, a third independent variable must be invoked.*

## The thermodynamic emissivity

It is clear at this point that if a PV system is to be subjected to the three governing laws of DB, EB, and SB, then these laws themselves have to be augmented to depend on a third independent variable. Our choice of this third variable is motivated by the fact that radiative recombinations are explicit functions of three quantifiers of a material system: the potential, the temperature, and the emissivity. Out of the three, only the emissivity has not yet been considered as an independent variable by either the DB or the thermodynamic approaches. We have also surveyed papers dealing with the measured emissivity of semiconductors, which showed that the sub-bandgap emissivity rises and the bandgap somewhat narrows at elevated temperatures[22–26]. Out of the two, the rise in emissivity seemed more pronounced. These observations led us to the following conclusion: The apparent emissivity of a material system has two major contributions. The first is the emissivity of the band-to-band (BTB) transitions, which is responsible for the creation of electron-hole pairs and their recombinations. The second is the abovementioned TE, which is the emissivity due to random fluctuations in a background thermal bath. The microscopic origin of the TE is irrelevant from the thermodynamic point of view, but it can be attributed to intra-valance or conduction band interactions or contributions from other parts of the electronic structure. The two emissivity terms compete since their sum is not allowed to surpass a value of unity on any part of the spectrum.

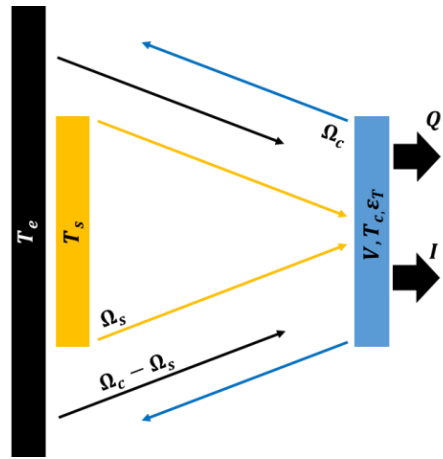

**Figure 1 | Schematic presentation of the system.** The system is characterized by a temperature $T_c$, potential $V$, and thermodynamic emissivity $\varepsilon_T$. The system radiates to its surroundings by emitting radiation at a solid-angle $\Omega_c$, and conducting heat at a rate $Q$ and an electrical current at a rate $I$. The system views a blackbody source of temperature $T_s$ at an angle of $\Omega_s$ and the ambient environment, which is at a temperature $T_e$, at an angle $\Omega_c - \Omega_s$.

To see how $\varepsilon_T$ affects the DB, EB, and SB laws, we look at Figure 1: A system characterized by $V$, $T_c$, and $\varepsilon_T$, exchanges radiation with a source at a temperature $T_s$ and with an environment at a temperature of $T_e$ ($T_e < T_s$). The viewing solid angle of the system is $\Omega_c$ and that of the source is $\Omega_s$, such that $\Omega_e = \Omega_c - \Omega_s \geq 0$. In addition, the system conducts electrical current $I$ to an external load and heat $Q$ to its environment. The purpose of the following analysis is to determine the independent variables $T$, $V$, and $\varepsilon_T$, for a given set of the above parameters. To do this, we assume an idealized form for the different emissivity terms, namely that $\varepsilon_T$ is uniform throughout the energy spectrum and that the BTB emissivity is zero below the system's bandgap, and $1 - \varepsilon_T$ above it. This choice of thermodynamic and BTB emissivity terms guarantees that their combined effect is $\varepsilon_T$ below the bandgap and one above it. Finally, we assume that the independent variables, $T$, $V$, and $\varepsilon_T$, are stationary and uniform throughout the system.

The DB law is only affected by the BTB processes and, therefore, is reformulated as follows:

$$(1 - \varepsilon_T)\left[\Omega_s N_g^\infty(0, T_s) + (\Omega_c - \Omega_s) N_g^\infty(0, T_e)\right] = (1 - \varepsilon_T)\Omega_c N_g^\infty(V, T_c) + I \quad (4)$$

The flux integrals are given in the following compact notation[28]:

$$N_a^b(V, T) \equiv \frac{2e}{c^2 h^3} \int_a^b \frac{E^2 \, dE}{\exp\left(\frac{E - V}{kT}\right) - 1},$$

where $e$, $c$, $h$, and $k$ are the electron charge ($eV$), the speed of light, Planck's constant, and Boltzmann's constant, respectively.

On the left of Eq. (4) are the source and environmental radiation fluxes that generate electron-hole pairs. On the right are the processes that eliminate electron-hole pairs from the system in the form of radiative BTB recombinations and electrical currents. According to our assumptions about the emissivity, all radiative BTB transitions are augmented by the required $(1 - \varepsilon_T)$ factor.

Next, the EB law now reads:

$$(1-\varepsilon_T)[\Omega_s E_g^\infty(0,T_s) + (\Omega_c - \Omega_s)E_g^\infty(0,T_e)] \quad (5)$$

$$+ \varepsilon_T[\Omega_s E_0^\infty(0,T_s) + (\Omega_c - \Omega_s)E_0^\infty(0,T_e)]$$

$$= (1-\varepsilon_T)\Omega_c E_g^\infty(V,T_c) + \varepsilon_T \Omega_c E_0^\infty(0,T_c) + E_{eh}I + Q$$

Also here, a compact notation for the energy flux integrals is adopted[28]:

$$E_a^b(V,T) \equiv \frac{2e}{c^2 h^3}\int_a^b \frac{E^3 dE}{exp\left(\frac{E-V}{kT}\right)-1}.$$

As in the DB law, the energy exchange with the BTB transitions is augmented by the required $(1-\varepsilon_T)$ factor. Unlike the DB law, however, the EB also includes an energy exchange with the thermal bath, which is augmented by the factor $\varepsilon_T$. The factors of $(1-\varepsilon_T)$ for the BTB processes and $\varepsilon_T$ for the thermal processes guarantee that on no occasion can the material absorb more than the entire energy flux of the sources.

As before, to the right are the energy depletion processes augmented by the same $(1-\varepsilon_T)$ and $\varepsilon_T$ factors for the BTB radiative recombinations of electrons, holes, and the radiation from the thermal bath, respectively. Here, $E_{eh}$ is the average energy of an electron-hole pair, such that $E_{eh}I$ is the energy carried by an electric current from the system of BTB excitations. Correspondingly, there is also the flux $Q$ of heat that is removed from the thermal bath, which can be conduction, convection, or even a blackbody radiation term.

Finally, there is the SB law:

$$S_g = (1-\varepsilon_T)S_g^\infty(V,T_c) + \varepsilon_T \Omega_c S_0^\infty(0,T_c) - (1-\varepsilon_T)[\Omega_s S_g^\infty(0,T_s) \quad (6)$$

$$+ (\Omega_c - \Omega_s)S_g^\infty(0,T_e)] - \varepsilon_T[\Omega_s S_0^\infty(0,T_s) + (\Omega_c - \Omega_s)S_0^\infty(0,T_e)]$$

$$+ S_{eh}I + S_Q \geq 0$$

The corresponding entropy flux integrals assume the following form[28,29]:

$$S_a^b(V,T) = \frac{2ke}{c^2h^3} \int_a^b E^2 \left[ \left(1 + \frac{1}{\exp\left(\frac{E-V}{kT}\right) - 1}\right) \ln\left(1 + \frac{1}{\exp\left(\frac{E-V}{kT}\right) - 1}\right) \right.$$

$$\left. - \frac{1}{\exp\left(\frac{E-V}{kT}\right) - 1} \ln\left(\frac{1}{\exp\left(\frac{E-V}{kT}\right) - 1}\right) \right] dE,$$

The SB law is identical to the EB law save for the inclusion of $S_g$: the entropy generation due to the generation and elimination of electron-hole pairs. Here $S_{eh}$ is the average entropy of an electron-hole pair such that $S_{eh}I$ is the entropy of the electric current and $S_Q$ is the entropy of the energy flux $Q$. This entropy flux is simply $\frac{Q}{T_c}$ if either convection or conduction is considered and is $\frac{4Q}{3T_c}$ for blackbody radiation.

As mentioned earlier, the optimal performance of the system, for a given set of external parameters, is attained from the minimization of $S_g$. Accordingly, the constraint imposed on the system by Eq. (6) is finding the set of independent variables that fulfill Eqs. (4) and (5), while minimizing $S_g$. An important property of the minimization process is observed by considering an electrically isolated and adiabatic system (isolated) under full concentration, where $I = Q = 0$ and $\Omega_c = \Omega_s$. It is easy to see that in in this case, the minimization of $S_g$ selects the *minimal* value of $V$ and the *maximal* value of $T_c$ among those that are allowed by the DB and EB of Eqs. (4) and (5). Since $V$ is the Gibbs free energy per electron-hole pair, the minimum of $S_g$ not only selects the $V$, $T_c$, and $\varepsilon_T$ that maximize the system's performance, but it also selects the system's minimal free energy and a temperature that equals that of the source. These are the only allowed values in this case since the system interacts with a single source and, thus, must be in a thermal equilibrium with it.

## Results & Discussion

We now examine a few cases of interest and discuss the unique features that arise due to the unification of the DB, EB, and SB laws. In order to facilitate a solution, the following straightforward definition for the average energy and entropy of an electron-hole pair are adopted:

$$E_{eh} = \frac{E_g^\infty(V,T_c)}{N_g^\infty(V,T_c)} \; ; \; S_{eh} = \frac{S_g^\infty(V,T_c)}{N_g^\infty(V,T_c)}. \tag{7}$$

Solving the model starts with finding the set of $0 \leq V \leq g$, $T_e \leq T_c \leq T_s$, and $0 \leq \varepsilon_T \leq 1$ values that fulfill Eqs. (4) and (5). These values are then fed into Eq. (6), and the unique state of the system is determined by the minimization of $S_g$.

### A completely isolated system

Let us first examine a completely isolated system, which is adiabatic and with an open circuit, where $I = Q = 0$. Figure 2 shows the temperature ($T_C$), thermodynamic emissivity ($\varepsilon_T$), and open circuit potential ($V_{oc}$) that emerges in this case as a function of the source concentration $C \equiv \frac{\Omega_s}{6.87 \times 10^{-5}}$ and for bandgaps of 1 ($eV$) and 1.42 ($eV$). Other parameters are $T_e = 300\ K$, $T_s = 5778\ K$, and $\Omega_c = \pi\ sr$. We see that $T_c$ and $\varepsilon_T$ rise with concentration, while $V_{oc}$ remains close to zero; the apparent fluctuations are attributed to numerical artifacts. At high concentrations, the system approaches the sun's temperature, regardless of its bandgap but with different values of $\varepsilon_T$, since it is an independent variable of the system. Also, $\varepsilon_T$ does not necessarily reach its maximal value of unity, even for its maximal concentration, since equilibrium between a selective emitter and a blackbody is possible. (Think of a semiconductor in a dark enclosure at room temperature; in this state of equilibrium, a step-like emissivity is expected.) The fact that $\varepsilon_T < 1$ is obtained for temperature beyond the melting point of any semiconductor is irrelevant for the case at hand since the bandgap narrowing effect was so far neglected.

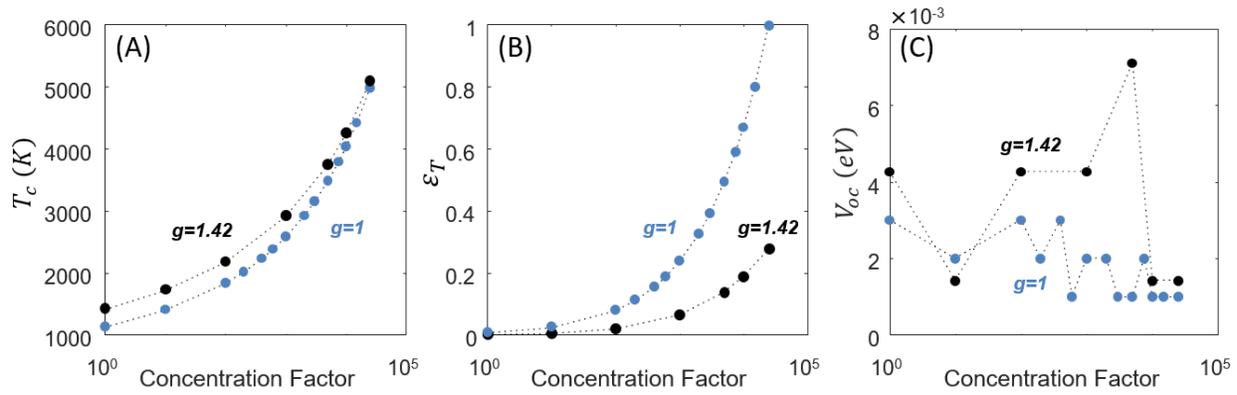

**Figure 2 | An isolated system**. The temperature (A), thermodynamic emissivity (B), and open circuit potential (C) for isolated systems ($I = Q = 0$) as a function of the sun's concentration. The results for two bandgaps of $g = 1\ (eV)$ and $g = 1.42\ (eV)$ are shown in black and blue, respectively.

## A system under open circuit conditions

Now, let us consider the open circuit case, where $I = 0$, but with heat conduction according to:

$$Q = \sigma(T - T_e), \qquad (8)$$

where $\sigma$ is the thermal conductivity, meaning that $S_Q = Q/T_c$. Figure 3 shows $T_c$, $\varepsilon_T$, and $V$ as a function of the source concentration $C$ at $g = 1.12\ eV$ and for different values of $\sigma$. All other parameters are the same as for the former isolated case. For comparison, we also show the results of the S&Q analysis of the same scenario, which is obtained by solving Eq. (4) alone for a fixed $T = 300\ K$.

We see that the system's temperature rises as it absorbs more energy and entropy from the source. Also, increased heat conduction reduces the system's temperature, as expected. We also see that for the lower concentrations, the temperature is asymptotic to that of the environment and for high concentrations to its value at $\sigma = 0$, as seen in Figure 1Figure 2. The transition between these two extremes is sensitive to the value of $\sigma$—the larger $\sigma$ is, the higher the concentration where this transition occurs. The thermodynamic emissivity is seen generally to rise with concentration but is zero until the temperature reaches its $\sigma = 0$ assymptot.

The significance of heat conduction is made clear by looking at the system's open circuit potential $V_{oc}$. Much like the temperature, $V_{oc}$ is asymptotic to its S&Q value at lower concentrations but eventually drops to zero as the flux increases, which is its $\sigma = 0$ limit. The point of departure from the S&Q value is sensitive to the value of $\sigma$—more heat conduction means the transition between asymptotes occurs at higher concentrations.

Finally, Figure 3 (B) and (C) shows an interesting feature of our model: The solution is either $\varepsilon_T = 0 \ \& \ V > 0$ or $\varepsilon_T > 0 \ \& \ V = 0$. This behavior is a direct result of the allowed values of $V$, $T_c$, and $\varepsilon_T$ according to Eqs. (4) and (5), and the entropy minimization process of Eq. (6). Such behavior was observed experimentally for the luminescence from rare earth metal[30].

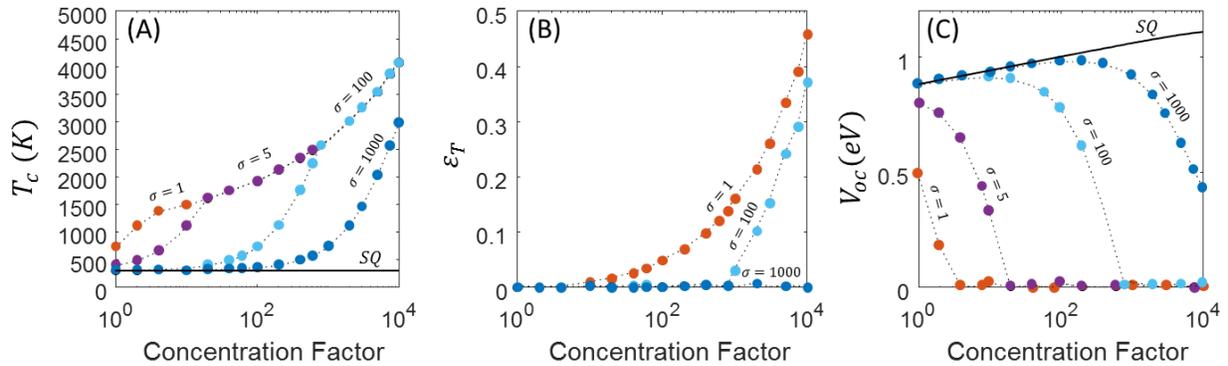

**Figure 3 | A system with heat conduction at open circuit**. The temperature (A), thermodynamic emissivity (B), and open circuit potential (C) as a function of the sun's concentration for different values of the heat conduction coefficient $\sigma$.

## A work producing system

Finally, we consider a system that is allowed to conduct heat, according to Eq. (8), and do work due to an electric current from some external load. Figure 4 shows the efficiency, in this case, calculated as $\max(I \cdot V)/[\Omega_s \cdot E_0^\infty(0, T_s)]$ as a function of the bandgap for different values of the heat conduction coefficient $\sigma$. The marker color on a specific heat conduction curve indicates the system's temperature according to the color bar on the right. The S&Q limit is calculated at an environmental temperature of $300K$ in this case.

We see that the efficiency approaches the S&Q limit for a larger heat conduction coefficient $\sigma$. For lower values, however, when $\sigma < 10$, for example, the efficiency deviates significantly from the prediction of S&Q. Expectedly, for a given value of $\sigma$, the temperature rises as the bandgap narrows, which is due to thermalization. Interestingly, the temperature reaches a maximum at a critical bandgap where the efficiency first nulls, owing to $V_{oc} = 0$. Beyond this point, an opposite trend appears where the temperature drops as the bandgap continues to narrow. This trend corresponds to a property of our model that favors a higher temperature for the smaller bandgap material as long as $V = 0$, which is also seen in Figure 2. Most importantly, however, Figure 4 shows that according to our model, the maximal efficiency is affected not only by the bandgap but also by the heat conduction from the system.

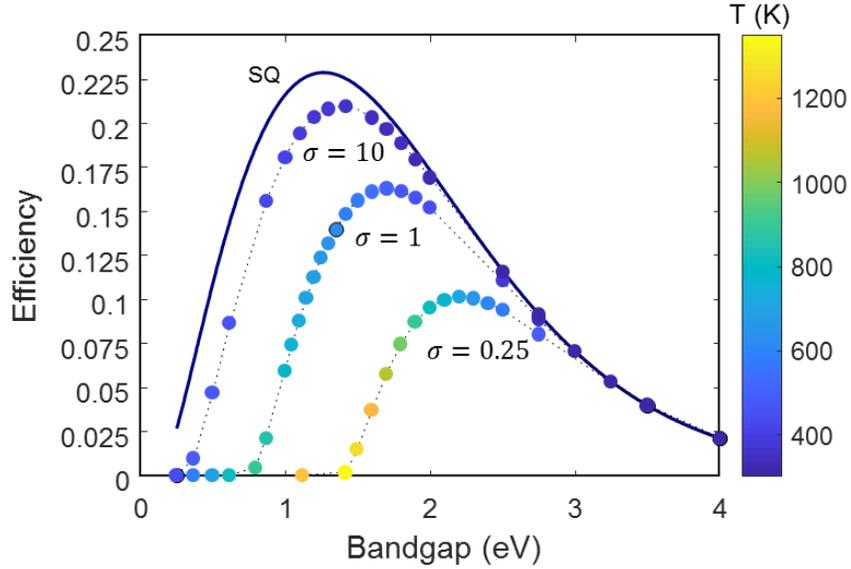

**Figure 4 | The efficiency as a function of the bandgap for different values of the thermal conductivity**. The figure shows different values of $\sigma$ ($0.25$, $1$, and $10$) for $T_e = 300\ K$, $T_S = 5778\ K$, $\Omega_s = 6.87 \times 10^{-5}$ (1 sun), and $\Omega_c = \pi$. The temperature is indicated by the color of the data points. The S&Q limit, marked by a solid black line, is taken for the same parameters and for a fixed $T_c = 300K$.

## Summary & Conclusions

This work addresses the inconsistency of the detailed balance approach with the laws of thermodynamics in their present form. The central observation is that a third independent variable must be considered. Based on the available parameters that control photon fluxes, and observation of semiconductors' emissivity at elevated temperatures, we propose that the missing variable is the emissivity of background thermal processes—the so-called thermodynamic emissivity. We show that the inclusion of this emissivity in the detailed balance, energy balance, and entropy balance laws provides a solvable system of constraints. The microscopic origin of the thermodynamic emissivity is attributed to electronic structural features that are distinct from the valance and conduction band-to-band transitions. The true nature of these processes is, however, irrelevant as far as the thermodynamic argument presented here is concerned.

Interestingly, our work points to a necessary connection between the band-to-band processes and the thermal background ones since the total emissivity may not surpass a value of one.

So far, we have considered only an idealized case. Non-radiative recombinations, however, such as Auger or Shockley-Read-Hall processes, can be readily incorporated into the model. Also, different environmental temperatures for the conduction and the radiative recombinations are allowed. In all the examples so far given, the source was taken as a blackbody at a temperature of 5778 K, which resembles the sun. This choice is not fundamental; any source with known photon, energy, and entropy fluxes can be used. Also, only a step emissivity function for the band to band processes was considered—a choice that entails a priori knowledge of the material bandgap. A generalization of this idealization that considers the finite absorptivity/emissivity of a real SC layer is possible, in a similar manner to what has been done with the S&Q model.[31–33] A more tangible aspect is the bandgap narrowing effect, the inclusion of which may cause radical changes to the model's predictions. Finally, the purpose of this article was only to present the physical argument behind the model and to illustrate it with a few examples. Any comprehensive mathematical survey of the model, which is needed, still remains out of scope.

While this new formalism may be of little value for terrestrial non-concentrating solar cells that work close to room temperatures, this is not the case for advanced PV schemes such as thermophotovoltaics[2], thermoradiative[3–5], thermophotonics[6], and extra-terrestrial concentrated multi-junction solar cells[11,12]. All of the above are designed to operate at elevated temperatures at which the thermodynamic emissivity can become a significant factor in their energy and entropy balances. The proposed approach is, therefore, crucial for the development of these potentially lucrative renewable energy technologies.

## Acknowledgments:

The authors would like to thank Prof. Jeffry Gordon for his thoughtful comments regarding the manuscript.

## Author Contributions

I. F. and A.N. developed the related theory, performed numerical calculations, and composed the manuscript. A.N. first recognized the inconsistency problem and its possible solution.

## Competing Interests

The authors declare no competing interest regarding the content of this article.

## Material & Correspondence

The custom code that supports the findings of this study is available from the corresponding author upon reasonable request. Corresponding author: Avi Niv (aviniv@bgu.ac.il).

## References:

1.  Shockley, W. & Queisser, J. H. Detailed Balance Limit of Efficiency of p-n Junction Solar Cells. *Am. Inst. Phys.* **32**, 510–519 (1961).

2.  Ferrari, C., Melino, F., Pinelli, M., Spina, P. R. & Venturini, M. Overview and status of thermophotovoltaic systems. *Energy Procedia* **45**, 160–169 (2014).

3.  Strandberg, R. Theoretical efficiency limits for thermoradiative energy conversion. *J. Appl. Phys.* **117**, (2015).

4.  Hsu, W. *et al.* Entropic and near-field improvements of thermoradiative cells. *Sci. Rep.* **6**, 34837 (2016).

5.  Zhou, Z., Sakr, E., Sun, Y. & Bermel, P. Solar thermophotovoltaics: Reshaping the solar spectrum. *Nanophotonics* **5**, 1–21 (2016).

6.  Harder, N. & Green, M. A. Thermophotonics. *Semicond. Sci. Technol.* **18**, S270–S278

(2003).

7. Zhu, L., Raman, A., Wang, K. X., Anoma, M. A. & Fan, S. Radiative cooling of solar cells. *Optica* **1**, 32–38 (2014).

8. Bermel, P., Boriskina, S. V., Yu, Z. & Joulain, K. Control of radiative processes for energy conversion and harvesting. *Opt. Express* **23**, A1533 (2015).

9. Sun, X., Sun, Y., Zhou, Z., Alam, M. A. & Bermel, P. Radiative sky cooling: Fundamental physics, materials, structures, and applications. *Nanophotonics* **6**, 997–1015 (2017).

10. Raman, A. P., Li, W., Raman, A. P., Li, W. & Fan, S. Generating Light from Darkness. *Joule* **3**, 1–8 (2019).

11. Pusch, A., Gordon, J. M., Mellor, A., Krich, J. J. & Ekins-Daukes, N. J. Fundamental Efficiency Bounds for the Conversion of a Radiative Heat Engine ' s Own Emission into Work. *Phys. Rev. Appl.* **12**, 064018 (2019).

12. Ruud, C. J. *et al.* Design and demonstration of ultra-compact microcell concentrating photovoltaics for space. *Opt. Express* **27**, A1467 (2019).

13. Landsberg, P. T. & Tonge, G. Thermodynamic energy conversion efficiencies. *J. Appl. Phys.* **51**, (1980).

14. De Vos, A. & Pauwels, H. On the thermodynamic limit of photovoltaic energy conversion. *Appl. Phys.* **25**, 119–125 (1981).

15. Baruch, P., De Vos, A., Landsberg, P. T. & Parrott, J. E. On some thermodynamic aspects of photovoltaic solar energy conversion. *Sol. Energy Mater. Sol. Cells* **36**, 201–222 (1995).

16. Landsberg, P. T. & Markvart, T. The Carnot factor in solar cell theory. *Solid. State. Electron.* **42**, 657–659 (1998).

17. Markvart, T. Thermodynamics of losses in photovoltaic conversion. *Appl. Phys. Lett.* **91**, 7–9 (2007).

(2003).

7. Zhu, L., Raman, A., Wang, K. X., Anoma, M. A. & Fan, S. Radiative cooling of solar cells. *Optica* **1**, 32–38 (2014).

8. Bermel, P., Boriskina, S. V., Yu, Z. & Joulain, K. Control of radiative processes for energy conversion and harvesting. *Opt. Express* **23**, A1533 (2015).

9. Sun, X., Sun, Y., Zhou, Z., Alam, M. A. & Bermel, P. Radiative sky cooling: Fundamental physics, materials, structures, and applications. *Nanophotonics* **6**, 997–1015 (2017).

10. Raman, A. P., Li, W., Raman, A. P., Li, W. & Fan, S. Generating Light from Darkness. *Joule* **3**, 1–8 (2019).

11. Pusch, A., Gordon, J. M., Mellor, A., Krich, J. J. & Ekins-Daukes, N. J. Fundamental Efficiency Bounds for the Conversion of a Radiative Heat Engine ' s Own Emission into Work. *Phys. Rev. Appl.* **12**, 064018 (2019).

12. Ruud, C. J. *et al.* Design and demonstration of ultra-compact microcell concentrating photovoltaics for space. *Opt. Express* **27**, A1467 (2019).

13. Landsberg, P. T. & Tonge, G. Thermodynamic energy conversion efficiencies. *J. Appl. Phys.* **51**, (1980).

14. De Vos, A. & Pauwels, H. On the thermodynamic limit of photovoltaic energy conversion. *Appl. Phys.* **25**, 119–125 (1981).

15. Baruch, P., De Vos, A., Landsberg, P. T. & Parrott, J. E. On some thermodynamic aspects of photovoltaic solar energy conversion. *Sol. Energy Mater. Sol. Cells* **36**, 201–222 (1995).

16. Landsberg, P. T. & Markvart, T. The Carnot factor in solar cell theory. *Solid. State. Electron.* **42**, 657–659 (1998).

17. Markvart, T. Thermodynamics of losses in photovoltaic conversion. *Appl. Phys. Lett.* **91**, 7–9 (2007).

18. Markvart, T. The thermodynamics of optical étendue. *J. Opt. A Pure Appl. Opt.* **10**, 15008 (2008).

19. Markvart, T. Solar cell as a heat engine: Energy-entropy analysis of photovoltaic conversion. *Phys. Status Solidi Appl. Mater. Sci.* **205**, 2752–2756 (2008).

20. Markvart, T. From steam engine to solar cells: can thermodynamics guide the development of future generations of photovoltaics? *Wiley Interdiscip. Rev. Energy Environ.* **5**, 543–569 (2016).

21. Ruppel, W. & Würfel, P. Upper Limit for the Conversion of Solar Energy. *IEEE Trans. Electron Devices* **27**, 877–882 (1980).

22. Sato, T. Spectral Emissivity of Silicon. **6**, 339–347 (1967).

23. Ravindra, N. M. *et al.* Temperature-dependent emissivity of silicon-related materials and structures. *IEEE Trans. Semicond. Manuf.* **11**, 30–39 (1998).

24. Timans, P. J. Emissivity of silicon at elevated temperatures. *Mater. Res. Soc. Symp. - Proc.* **525**, 27–38 (1998).

25. Moss, T. S. & Hawkins, T. D. H. The infra-red emissivities of indium antimonide and germanium. *Proc. Phys. Soc.* **72**, 270–273 (1958).

26. Timans, P. J. The experimental determination of the temperature dependence of the total emissivity of GaAs using a new temperature measurement technique. *J. Appl. Phys.* **72**, 660–670 (1992).

27. Landsberg, P. T. & Tonge, G. Thermodynamics of the conversion of diluted radiation. *J. Phys. A. Math. Gen.* **12**, 551 (1979).

28. Würfel, P. & Ruppel, W. The chemical potential of luminescent radiation. *J. Lumin.* **24–25**, 925–928 (1981).

29. L D Landau E.M. Lifshitz. *Statistical Physics 3rd Edition*. (Elsevier, 2013).

30. Manor, A., Martin, L. & Rotschild, C. Conservation of photon rate in endothermic

photoluminescence and its transition to thermal emission. *Optica* **2**, 0–3 (2015).

31. Tiedje, T., Yablonovitch, E., Cody, G. D. & Brooks, B. G. Limiting Efficiency of Silicon Solar Cells. *IEEE Trans. Electron Devices* **31**, 711–716 (1984).

32. Yu, Z., Raman, A. & Fan, S. Fundamental limit of nanophotonic light trapping in solar cells. *Proc. Natl. Acad. Sci. U. S. A.* **107**, 17491–6 (2010).

33. Niv, A., Gharghi, M., Gladden, C., Miller, O. D. & Zhang, X. Near-field electromagnetic theory for thin solar cells. *Phys. Rev. Lett.* **109**, 1–13 (2012).